\documentclass{revtex4}
\usepackage{graphicx}
\usepackage{latexsym}
\usepackage{amsmath,amssymb}
\usepackage{epsfig,graphicx}
\def\be{\begin{equation}}
\def\ee{\end{equation}}
\def\bea{\begin{eqnarray}}
\def\eea{\end{eqnarray}}
\begin{document}

%\begin{flushright}
%\today\\
%\end{flushright}

\title{The cosmological constant and dark energy in braneworlds}
\author{Kazuya Koyama}
\email{kazuya.koyama.AT.port.ac.uk}%
\affiliation{Institute of Cosmology \& Gravitation, University of
Portsmouth, Portsmouth~PO1~2EG, UK}

\begin{abstract}
We review recent attempts to address the cosmological constant 
problem and the late-time acceleration of the Universe based on 
braneworld models. In braneworld models, the way in which the vacuum 
energy gravitates in the 4D spacetime is radically different
from conventional 4D physics. It is possible that the vacuum 
energy on a brane does not curve the 4D spacetime and only affects 
the geometry of the extra-dimensions, offering a solution 
to the cosmological constant problem. We review the idea of 
supersymmetric large extra dimensions that could achieve this 
and also provide a natural candidate for a quintessence 
field. We also review the attempts to explain the late-time 
accelerated expansion of the universe from the 
large-distance modification of gravity based on the 
braneworld. We use the Dvali-Gabadadze-Porrati model 
to demonstrate how one can distinguish this model from
dark energy models in 4D general relativity. Theoretical 
difficulties in this approach are also addressed. 
\end{abstract}

\maketitle

\section{Introduction}
The cosmological constant problem is a long-standing problem in physics
\cite{Wei}.
Particle physics predicts the existence of the vacuum energy density 
which is related to the fundamental scale of the theory, like the electroweak 
scale, $\rho_{vac} \sim (\mbox{TeV})^4$. This is typically more than 
50 orders of magnitude larger than the observed value, $\rho_{\Lambda} \sim (10^{-3} \mbox{eV})^4$. 
Before the discovery of the accelerated expansion 
of the Universe, physicists tried to answer this question by seeking 
a theory that predicts the cosmological constant {\it should be} zero. 
However, the discovery of the accelerated expansion of 
the Universe makes this answer insufficient \cite{Sch, Gar0, Per}. Now, we should explain why it is 
non-zero and yet it is so small. Moreover, there is a coincidence problem.
The cosmological constant dominates the energy density of the Universe 
only recently. If the cosmological constant is really a constant, 
we should explain {\it why now}, does it become dominant. 

One direction to answer these questions is to appeal to the anthropic 
principle \cite{Wei}. If the cosmological constant is too large, the 
accelerated expansion started too early and it prevents structure
from growing and we cannot exist. On the other hand, a universe with 
negative comsological constant re-collapses. Then observers will 
only exist within a tiny anthropic range of cosmological constant
(see for example \cite{Gar}).
This idea is strengthened by the discovery in string 
theory that there are millions of low-energy vacua in the theory 
(the string theory landscape) \cite{Sus}. It is argued that we might need 
the anthropic principle to select the low-energy vacuum. 
However, many theorists still hope
to explain the problem without invoking the existence of ourselves
in the Universe. Although significant efforts have been devoted to 
this attempt, we still have not succeeded yet to provide convincing models. 
However, the rapid progress of string theory has provided a new perspective 
for solving this problem. In this review, we focus on the attempts 
of using higher-dimensional gravity and branes to address the 
problem. 

String theory is formulated in a 10D spacetime. On the other hand, 
our observed Universe is a 4D spacetime. Thus there should be a 
mechanism to hide the extra dimensions. The conventional idea is to compactify 
the extra dimensions by the Kaluza-Klein (KK) mechanism. The size of 
the extra dimensions $L$ should be small, $L < \mbox{TeV}^{-1}$,
in order not to spoil the success of the standard model 
of particle physics that is formulated in a 4D spacetime. 
Below the energy scale determined by the size of the
extra dimensions, $L^{-1}$, the universe looks completely 4D 
if the radius of the extra dimensions
is stabilized. Recently, a completely new way of hiding the extra dimensions
has been proposed. This is the brane world mechanism where 
matter fields are confined to a 4D membrane in a higher dimensional 
spacetime (see \cite{Mar1} for a review). 
Only gravity and non-standard model particles can propagate 
into the whole higher-dimensional bulk spacetime. In this picture, 
the size of the extra dimensions can be much larger than that in the conventional 
KK compactification. In fact, the size of the extra dimensions could even be 
infinite. If the bulk is a spacetime with a negative cosmological 
constant, that is, an Anti-de Sitter (AdS) spacetime, it is shown 
that gravity behaves like 4D on scales larger than the 
AdS curvature length, even if the size of the extra dimensions is infinite \cite{Ran}. 
Another way is to introduce induced gravity
on a brane \cite{Dva} (see \cite{Aka0} for an early attempt). 
If we assume there is an Einstein-Hilbert term on a
brane, 4D gravity is recovered, in this case, on small scales 
even if the bulk is an infinite Minkowski spacetime. In these 
braneworld models, the behaviour of gravity can be dramatically 
different from the 4D theory, providing a new perspective to 
solve the cosmological constant and the dark energy problem.

This article will review several approaches to address the 
cosmological constant and the late-time acceleration problem
based on braneworld gravity. Firstly, we explain 
the attempts to address the `old' cosmological constant
problem $-$~why the vacuum energy is incredibly small compared 
with the prediction of particle physics. These attempts 
exploit the modification of 4D gravity in the braneworld 
and change the way in which the vacuum energy gravitates in 
a 4D spacetime.
Secondly, we introduce an idea to explain the late-time 
acceleration without introducing the cosmological 
constant. This idea also relies on the modification of 
gravity on large scales based on the braneworld idea.

In section II, we give a brief introduction to braneworlds.
In section III, the attempts to solve the old cosmological 
constant problem are discussed. In section IV, the idea to 
realize late-time acceleration without introducing a 
cosmological constant is explained. 
Section V is devoted to conclusions.

\section{Braneworld models}
The idea that ordinary matter fields are 
confined to a lower-dimensional domain wall 
was proposed in the 1980's \cite{Aka, Rub}. It was shown that 
fermion fields can be confined to a field 
theoretic domain wall. The progress in string 
theory, especially the discovery of D-branes, has revived 
these attempts \cite{Pol}. The D-brane is defined by a membrane
on which end-points of open strings lie. 
At the end-points of open strings, gauge fields can be 
attached. Then gauge fields are confined to the D-brane. 
On the other hand, closed strings that contain the graviton 
can propagate into the whole bulk.  Then there arises 
a braneworld picture where usual matter fields 
are confined to a brane while gravity propagates 
into the whole bulk spacetime. 
A schematic picture of the braneworld 
is shown in Fig.~1. Based on this idea, 
several simplified braneworld models have been 
proposed that capture the basic features of 
the braneworld, yet in which we can address 
many important problems from a new perspective. 

\begin{figure}[h]
  \begin{center}
 \includegraphics[width=8cm]{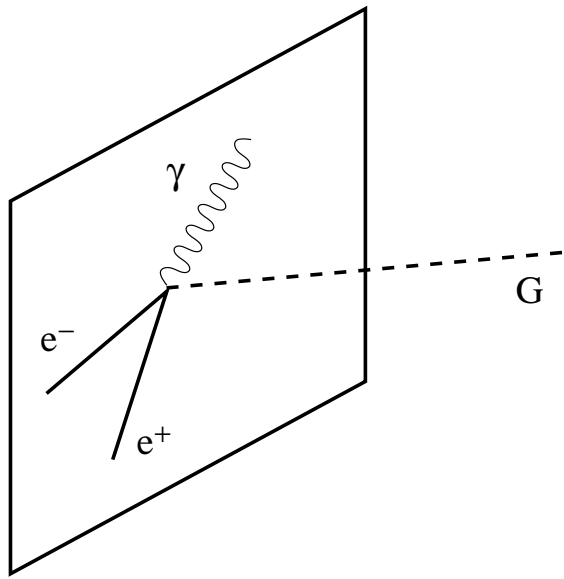}
  \end{center}
  \caption{A schematic picture of the braneworld. From 
  \cite{Mar1}.}
  \label{ecfig}
\end{figure}
\subsection{Arkani-Hamed-Dimopoulos-Dvali model}
An interesting possibility in braneworld models 
is that some of the extra dimensions can be {\it large} 
\cite{Ark0, Ant, Ark}.
In a conventional picture, extra dimensions are 
rolled up small so that we never observe them. 
More precisely, in order not to spoil the success 
of the standard model of particle physics that is 
formulated in a 4D spacetime, the size of the extra dimensions 
should be smaller than TeV$^{-1} \sim 10^{-19}$ m. However, 
in the braneworld, the standard model particles 
are confined to the 4D brane. Thus we do not need 
to worry about this constraint. The gravitational 
interactions are very weak and the 4D behaviour of 
the Newtonian force is  
only verified down to $44\mu$m \cite{Kap}. Thus the size of the 
extra-dimensions is allowed to be as large as $44\mu$m.

\begin{figure}[!]
\hfil\scalebox{.55}{\includegraphics*[64pt,29pt][510pt,431pt]{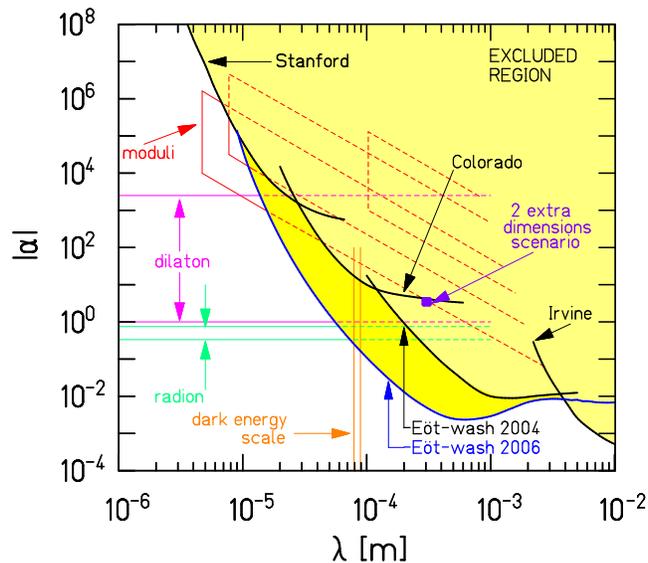}}\hfil
\caption{Constraints on Yukawa violations of the gravitational $1/r$ potential
, $V(r) \propto (1/r) (1 + \alpha \exp (-r/\lambda))$.
The shaded region is excluded at the 95\% confidence level. From \cite{Kap}.}
\label{fig: constraints}
\end{figure}

This opens up a new perspective to solve another 
serious problem in particle physics, namely the 
hierarchy problem: why the gravitational interaction 
is so weak compared with the other interactions. The 
answer could be that the gravitational field of 
an object on a brane leaks out into the large extra-dimensions 
and this leakage weakens the gravitational interactions on 
a brane. The gravitational potential generated by an object 
with mass $M$ is given by 
\begin{eqnarray}
\Psi(r) &=& - \frac{G_4 M}{r^2}, \quad (r > L), \\
\Psi(r) &=& - \frac{G_D M}{r^{D-2}}  \quad (r < L),
\end{eqnarray}
where $L$ is the size of the $(D-4)$ dimensional 
extra-dimensions. Then the 4D  
gravitational constant is given in terms of 
the higher-dimensional gravitational constant as
\begin{equation}
G_4 = \frac{G_D}{L^{D-4}}, \quad M_4 = M_D^{(D-2)/2} L^{(D-4)/2},
\label{newton}
\end{equation}
where $8 \pi G_D = M_D^{-(D-2)}$. 
Then even if the fundamental scale of gravity $M_D$ 
is TeV, the 4D gravitational constant can be $10^{19}$ GeV
as long as $L$ is appropriately large. For example, for $D =6$,
the current constraint on the deviation from the 
gravitational inverse-square law $L < 44\mu$m implies 
$M_6 >3.2$ TeV.

\subsection{Randall-Sundrum model}
The most difficult problem in the braneworld 
in terms of gravity is the inclusion of the self-gravity of the branes. 
In the ADD model, the self-gravity of the branes is implicitly neglected.
The model proposed by Randall and Sundrum (RS) offers a 
consistent framework to deal with higher-dimensional 
gravity including the self-gravity of the branes \cite{Ran}. 
They consider a 5D spacetime described by the action 
\begin{equation}
S = \frac{1}{2 \kappa_5^2} 
\int d^5 x \sqrt{-g} ({}^{(5)}R - 2 \Lambda) 
- \sigma \int d^4 x \sqrt{-\gamma} 
+ \int d^4 x {\cal L}_m,
\end{equation}
where $\kappa_5^2 = 8 \pi G_5$ and ${\cal L}_m$ represents the matter 
lagrangian confined to a brane.
The introduction of the singular objects enforces the 
junction condition (Israel junction condition) 
at the location of the brane. The junction condition 
relates the extrinsic curvature at the brane to the energy 
momentum tensor localized on a brane. 
By solving the 5D bulk spacetime and imposing the junction 
condition at the brane, the solution for the gravitational field 
on the brane is obtained. 
The simplest solution is a solution with a Minkowski brane.
The 5D metric is given by
\begin{equation}
ds^2 = dy^2 + \exp(-2 |y|/\ell) \eta_{\mu \nu} dx^{\mu} dx^{\nu}.
\end{equation}
A brane is located at $y=0$ and the reflection symmetry ($Z_2$ 
symmetry) across the brane is imposed. The exponential `warp 
factor' is an essential ingredient of the model. Even if 
the physical size of the fifth dimension is infinite, 
low-momentum gravity is confined near the brane due to 
the curvature of the bulk spacetime and 4D gravity 
is recovered. It is shown that the solutions for weak gravity at large 
distances $r \gg \ell$ are given by \cite{Gar2}
\begin{equation}
ds^2 = -(1 +2 \Psi) dt^2+ (1+ \Phi) \delta_{ij} dx^i dx^j,
\end{equation}
\begin{equation}
\Psi= \frac{2G_4 M}{r} \left(1 + \frac{2 \ell^2}{3 r^2} \right), \quad
\Phi = \frac{2G_4 M}{r} \left(1 + \frac{\ell^2}{3 r^2} \right),
\end{equation}
where $\kappa_4^2 = 8 \pi G_4$ and is determined by
\begin{equation}
G_4 = G_5 \ell.
\end{equation}
Comparing this with Eq.~(\ref{newton}), we notice that $\ell$ acts as the 
effective size of the extra dimension. Thus the 
RS model provides an `alternative to compactification'. 

Despite the remarkably simple setup of the model, gravity 
in this model is incredibly complicated. Fortunately, 
for a homogeneous and isotropic brane, the generalized 
Birkoff theorem ensures that the bulk spacetime is AdS spacetime 
or AdS-Schwarzchild spacetime. Then the Friedmann equation on the 
brane is easily derived as \cite{Bin, Ida, Kra}
\begin{equation}
H^2 = \frac{\Lambda_4}{3} + \frac{\kappa_4^2}{3}\rho + 
\frac{\kappa_5^4}{36} \rho^2 + \frac{C}{a^4},
\end{equation}
where
\begin{equation}
\Lambda_4 = \frac{\Lambda_5}{2} + \frac{\kappa_5^4 \sigma^2}{12},
\quad \kappa_4^2 =\frac{\kappa_5^2 \sigma}{6}.
\label{lambdaRS}
\end{equation}
The constant $C$ is proportional to the black hole mass in 
the bulk. In accord with weak gravity, cosmology also shows the transition 
from 4D to 5D. At high energies $H \ell >1$ where the horizon size 
$H^{-1}$ is smaller than the effective size of the extra-dimension $\ell$, 
the Friedmann equation is significantly modified and $H \propto \rho$. 
At low energies $H \ell <1$, we recover the 4D Friedmann equation. 

\section{Cosmological constant problem in the braneworld}

\subsection{Self-tuning 5D braneworld}
The relation between the vacuum energy and the effective 
cosmological constant on a brane is different from 
that in the usual 4D theory. In the RS braneworld, the 
vacuum energy in the brane $\sigma$ is not 
directly related to the cosmological constant $\Lambda_4$ on 
the brane in the effective Einstein equation as in Eq.~(\ref{lambdaRS}).
In the RS braneworld, there should be a cancellation between 
the 4D and 5D contribution of the vacuum energy in order to 
have a vanishing cosmological constant on the brane. This 
requires a fine-tuning for the parameters in the action. 
Instead of having the cosmological 
constant in the bulk and tension on the brane, 
let us consider a scalar field with potentials \cite{Ark2, Kac}.
The action is given by
\begin{equation}
S = \frac{1}{2 \kappa_5^2} \int d^5 x \sqrt{-g} \left(   
R - \frac{4}{3} (\partial_{\mu} \phi)^2 - V(\phi)
\right) - \int d^4 x \sqrt{-\gamma} f(\phi).
\end{equation}
The potentials can be taken as 
\begin{equation}
V(\phi) =\Lambda_0 \exp(a \phi), \quad f(\phi)=
V_0 \exp(b \phi).
\end{equation}
With this choice, the action describes a family 
of theories parametrized by $V_0, \Lambda_0, a$ and $b$.
For simplicity, we take $\Lambda_0=0$. We look for 
a solution with a Minkowski spacetime on a brane. 
The 5D metric is given by
\begin{equation}
ds^2 = dy^2 + e^{2 A(y)} \eta_{\mu \nu} dx^{\nu} dx^{\mu}.
\end{equation}
The 5D Einstein equation gives the relation between 
the warp factor $A(y)$ and the scalar field $\phi(y)$
\begin{equation}
\phi'(y) = \pm \frac{1}{3} A'(y).
\end{equation}
The solution for $\phi$ in the bulk is then obtained as
\begin{eqnarray}
\phi(y) &=& \frac{3}{4} \log \left( 
\frac{4}{3} M_5 y +c_1 \right) + d_1, \quad y <0, \\
\phi(y) &=& -\frac{3}{4} \log \left(  
\frac{4}{3} M_5 y +c_2 \right) + d_2, \quad y >0,
\end{eqnarray}
where $c_1, c_2, d_1$ and $d_2$ are integration constants.
The continuity of $\phi$ determines $d_2$.
Then the junction conditions for the scalar field 
and the warp factor determine $c_1$ and $c_2$ 
in terms of $b, V_0$ and $d_1$ if $b \neq \pm 4/3$.
This means that for a scalar coupling given by $b$, 
there is a Minkowski solution on a 4D brane for 
{\it any} value of the brane tension $V_0$. This is the 
idea of the `self-tuning'. The vacuum energy in 
a 4D brane is cancelled by the integration constants
in the solutions, not by the parameters in the original
action. Thus this is not a fine-tuning. 
The hope is that the solution in the bulk adjusts itself 
so that the contribution from the vacuum energy on the 
brane is exactly cancelled. 

Although the idea of self-tuning is very attractive, 
there are several problems in the original proposal
\cite{For, For1, Csa, Bin2}. 
Firstly, there is a naked singularity in the above 
model with a scalar field. Any procedure that regularizes
the singularity in the solutions would cause the re-introduction
of the fine tuning. There is also a problem of stability.
In the case of vanishing potential in the bulk, 
the static solution is unstable, leading to a singularity. 
A modified version of the model using the bulk 
black hole to hide the singularity inside the 
horizon was proposed \cite{Csa2}, but it was argued 
that this model also cannot avoid the fine tuning \cite{Cli}. 

\subsection{6D braneworld}
Another approach to realize the self-tuning is  
to consider a 6D bulk spacetime \cite{Car}. 
The action is given by 
\begin{equation}
S = \int d^6 x \sqrt{-g} 
\left (\frac{1}{2 \kappa_6^2} R - \Lambda_6 - \frac{1}{4}
F_{ab}F^{ab} \right),
\end{equation}
where the gauge field $F_{ab}$ is required to stabilize
the size of the extra dimensions. We decompose the coordinates into four
macroscopic dimensions and the two extra dimensions.
The metric is taken as 
\begin{equation}
ds^2  = \eta_{\mu \nu} dx^{\mu}dx^{\nu} + \gamma_{ij}
dx^{i} dx^{j}.
\end{equation}
The gauge field is taken to consist of magnetic flux threading 
the extra dimensional space so that the field strength 
takes the form 
\begin{equation}
F_{ij} = \sqrt{\gamma} B_0 \epsilon_{ij},
\end{equation}
where $B_0$ is a constant, $\gamma$ is the determinant of $\gamma_{ij}$
and $\epsilon_{ij}$ is the antisymmetric tensor normalized as 
$\epsilon_{12} =1$. All other components of $F_{ab}$ vanish. 
A static and stable solution is obtained by choosing the 
extra-dimensional space to be a two-sphere 
\begin{equation}
\gamma_{ij} dx^{i} dx^{j} =a_0^2 (d \theta^2 + \sin^2 \theta d \varphi^2).
\end{equation}
The magnetic field strength $B_0$ and the radius $a_0$ are fixed by 
the cosmological constant 
\begin{equation}
B_0^2 = 2 \Lambda_6, \quad a_0^2 = \frac{M_6^4}{2 \Lambda_6}.
\label{tuning}
\end{equation}
It should be noted that $B_0$ has to be tuned so that a Minkowski 
spacetime is induced in 4D. 
Now we add branes to this solution. The brane action is 
given by 
\begin{equation}
S_4 = - \sigma \int d^4 x \sqrt{-\gamma}.
\end{equation}
The solution for the extra dimensions is now given by
\begin{equation}
\gamma_{ij} dx^i dx^j = a_0^2 (d \theta^2 + \alpha^2 
\sin^2 \theta d \varphi^2),
\end{equation}
where
\begin{equation}
\alpha = 1 - \frac{\sigma}{2 \pi M_6^2}, \quad 
a_0^2 =\frac{M_6^4}{2 \Lambda_6}.
\end{equation}
The coordinate $\varphi$ ranges from $0$ to $2 \pi$. Thus the 
effect of the brane makes a deficit angle 
$\delta = 2 \pi(1- \alpha)$ in the bulk. 
This is a 6D realization of the ADD model 
including the self-gravity of branes.

\begin{figure}
  \begin{center}
 \includegraphics[scale=1]{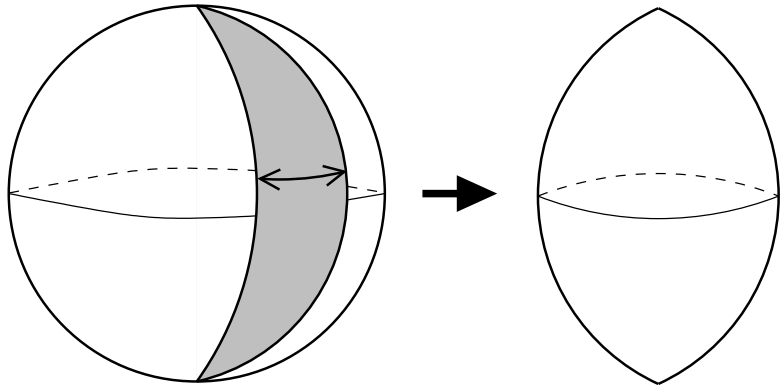}
  \end{center}
  \caption{Removing a wedge from a sphere and identifying opposite
  sides to obtain a football geometry.  Two equal-tension branes with
  conical deficit angles are located at either pole; outside the
  branes there is constant spherical curvature. From 
  \cite{Car}.}
  \label{6d}
\end{figure}

The most interesting feature of this solution is that
the 4D geometry is independent of the brane tension $\sigma$.
The tension enters only in the deficit angle and not 
the radius $a_0$ and the magnetic field $B_0$ that need
to be tuned to obtain a flat 4D spacetime. Thus the vacuum 
energy on the brane does not gravitate in the 4D spacetime
but merely changes the geometry of the extra dimensions. 
Thus the outcome is similar to the self-tuning solutions 
discussed in section III.A. 
It should be noted that the cosmological constant problem 
is not fully solved even if this idea works. In order to 
obtain a flat spacetime, we need to tune the magnetic field 
and the bulk cosmological constant as in Eq.~(\ref{tuning}). 
However one could hope that some kinds of symmetry like supersymmetry 
in the bulk can ensure this tuning. 

However, there have been objections to the self-tuning in this 
model \cite{Gar3}. Consider that a phase transition occurs and the tension 
of the brane changes from $\sigma_1$ to $\sigma_2$. Accordingly, 
$\alpha$ changes from $\alpha_1 =1-\sigma_1/(2 \pi M^4)$ to 
$\alpha_2 = 1-\sigma_2/(2 \pi M^4)$. The magnetic flux is 
conserved as the gauge field strength is a closed form,  
$d F=0$. Then the magnetic flux which is obtained 
by integrating the field strength over the extra dimensions
should be conserved
\begin{equation}
\Phi_B= 4 \pi \alpha_1 B_{0,1} = 4 \pi \alpha_2 B_{0,2}.
\end{equation}
The fine-tuning of $\Lambda_6$ and $B_0$, Eq.~(\ref{tuning}), that ensures 
the existence of Minkowski branes cannot be imposed both 
for $B_0 = B_{0,1}$ and $B_{0}=B_{0,2}$ when $\alpha_1 \neq
\alpha_2$. This becomes clear if we rewrite 
the conditions Eq.~(\ref{tuning}) as
\begin{equation}
\alpha^2 = \left( \frac{\Phi_B}{4 \pi} \right)^2 
\frac{\Lambda_6}{M_6^4}.
\end{equation}
The left-hand side changes by the phase transition
but the right-hand side cannot change. Moreover, 
the quantization condition must be imposed on 
the flux $\Phi_B$. Then if the condition Eq.~(\ref{tuning})
is satisfied for some value of $\alpha$, it will not 
be satisfied by neighbouring values. 
Thus after the phase transition, the 4D spacetime cannot 
be static \cite{Vin1}. 

\subsection{Supersymmetric large extra dimensions}
In the Einstein-Maxwell theory discussed in 
section III.B, the tuning between the 
magnetic flux and the cosmological constant in 
the 6D spacetime, Eq.~(\ref{tuning}), was necessary 
to obtain the flat 4D spacetime. This was 
the origin of the difficulty in realizing the self-tuning.
To evade this problem, the Supersymmetric Large 
Extra Dimensions (SLED) model was proposed (see \cite{Bur1, Bur2} 
for a review). 
This is a supersymmetric version of the 6D model and the 
action is given by \cite{Agh}
\begin{equation}
S =\int d^6 x \sqrt{-g}
\left[ \frac{1}{2 \kappa_6^2}
\left(R -\partial_M \phi \partial^M \phi \right)
-\frac{1}{4} e^{-\phi} F_{MN}F^{MN}- e^{\phi} \Lambda_6 
\right].
\end{equation}   
There exists a solution where the dilaton $\phi$ 
is constant, $\phi=\phi_0$, and the solution in 
the Maxwell-Einstein theory is a solution just 
by replacing $\Lambda_6 \to \Lambda_6 e^{\phi_0}$
and $B_0^2 \to B_0^2 e^{-\phi_0}$. The constant 
value $\phi_0$ is determined by the condition 
that the potential for $\phi$ has minimum \cite{Vin2}
\begin{equation}
V'(\phi_0) = -\frac{1}{2} B_0^2 e^{-\phi_0}
+\Lambda_6 e^{\phi_0} =0.
\end{equation}
This is exactly the condition to have a flat geometry
on the brane (see Eq.~(\ref{tuning}))
\begin{equation}
B_0^2 e^{-\phi_0}= 2 \Lambda_6 e^{\phi_0}.
\label{finetune}
\end{equation}
Thus unlike the Einstein-Maxwell system, one might 
not need a tuning condition in the bulk. 
In fact, The known solutions in this model
which have maximally symmetric 4D metric all have 
vanishing vacuum energy. 

Again there were several objections to this version of 
the self-tuning \cite{Gar3, Vin2}. It is possible to derive the 
4D effective theory by putting the metric 
in the form
\begin{equation}
ds^2 = g_{\mu \nu}(x) dx^{\mu} dx^{\nu}
+M_6^{-2} e^{-2\psi(x)} (dr^2+ \sin^2 r d\theta^2),
\label{ansatz}
\end{equation}
and assuming $\phi= \phi(x)$. The potential which results 
from the two scalar fields is \cite{Gar3}
\begin{equation}
V(\psi,\phi) =M_6^{-4} e^{\sigma_2} U(\sigma_1),
\quad U(\sigma_1)= 
\frac{B_0^2}{2 \alpha^2} e^{-2 \sigma_1} 
-2 M_6^2 e^{-\sigma_1} + 2 \Lambda_6,
\end{equation}
where $\sigma_1 =2 \psi + \phi$ and $\sigma_2 = 2 \psi -\phi$.
Unlike the Einstein-Maxwell theory, $\sigma_2$ 
ensures that $U(\sigma_1)$ vanishes at the 
minimum of the potential. We should note that $\sigma_2$
is related to the classical scaling property of the model.
The 6D equation of motion is invariant under the constant
rescaling $g_{MN} \to e^{\omega} g_{MN}$ and $e^{\phi}
\to e^{\phi -\omega}$ and the lagrangian is scaled as 
${\cal L} \to e^{2 \omega} {\cal L}$. The modulus $\sigma_2$
can be identified as the one associated with this scaling 
property. Thus the flatness of the 4D spacetime is 
ensured by the scaling property of the theory.  
However, this eventually leads to the same tuning 
condition (\ref{finetune}) as in the Einstein-Maxwell theory. 
Then we can apply the same arguments as in the previous 
section. Suppose that the tension of the brane changes.
Flux conservation (and flux quantization) 
means that the tuning condition cannot be maintained and 
$U(\sigma_1) \neq 0$. What happens would be that $\sigma_2$ 
acquires a runaway potential and the 4D spacetime 
becomes non-static. 

A caveat in this argument is that the metric 
ansatz (\ref{ansatz}) is restrictive. In fact, 
there is a class of static solutions where there is a warping in 
the bulk. The solution has the form \cite{Gib, Agh2}
\begin{equation}
ds_6^2 = W(\eta)^2 \eta_{\mu \nu}dx^{\mu} dx^{\nu} 
+ a_0^2 (W(\eta)^8 d \eta^2 + d \theta^2),
\quad \phi = \phi_0 + 4 \ln W(\eta) + 2 \lambda_3 \eta,
\end{equation} 
where $W(\eta)$ is the warp factor. If both branes, 
at the north pole and the south pole, have the same tension,
the warp factor becomes trivial. However, if the 
tensions are not equal, there is a warping. 
For $\lambda_3 \neq 0$, the metric near the branes no longer corresponds 
to that of a simple conical singularity. These solutions cannot 
be described by the ansatz (\ref{ansatz}). Thus one can still 
hope that the solutions will go to these solutions after a 
change of tension. An unambiguous way to investigate this 
problem is to study the dynamical solutions directly in the 
6D spacetime. However,
once we consider the case where the tension becomes time 
dependent, we encounter a difficulty to deal with the 
branes. This is because for co-dimension 2 branes, we 
encounter a divergence of metric near the brane if 
we put matter other than tension on a brane. 
Hence, without specifying how we regularize the branes, 
we cannot address the question what will happen if we change 
the tension. Is the self-tuning mechanism at work and does it lead 
to a 4D static solution? 
Or do we get a dynamical solution driven by the runway 
behaviour of the moduli field? 
There was a negative conclusion on the self-tuning in this 
supersymmetric model for a particular kind of regularization 
\cite{Vin2}. However, the answer could depend on the regularization 
of branes and the jury remains out. It is important to 
study the time-dependent dynamics in the 6D spacetime
and the regularization of the branes 
\cite{Bur3, Bur4, Kob, Cop}.

If the self-tuning mechanism works, then we should 
seek an explanation for the accelerated expansion today. 
The supersymmetry in the bulk would also provide a 
very interesting mechanism (see \cite{Bur1, Bur2} 
and references therein for detailed discussions). Supersymmetry is 
supposed to be broken at least at the electroweak scale
$M_w$. Then in the 4D spacetime, this gives a vacuum 
energy of the order $\rho \sim M_w^4$ as the cancellation 
between the contribution to the vacuum energy from boson fields 
and fermion fields ceases to exist at $M_w$. However, if 
the self-tuning mechanism is at work, this vacuum energy 
does not give any contribution to the cosmological 
constant on the brane. However, the breakdown of 
supersymmetry is mediated to the bulk at least gravitationally. 
Then there arises a supersymmetry breaking scale 
in the bulk given by 
\begin{equation}
M_{sb} = \frac{M_w^2}{M_{4}}.
\end{equation} 
Interestingly, this scale is related to the size of the extra dimensions.
If we want to solve the hierarchy problem between the Planck 
scale and the electroweak scale, $M_6$ should be of the order 
$M_w$. Then from the relation between $M_6$, $M_{4}$ and the size 
of the extra-dimension $L$, Eq.~(\ref{newton}), the supersymmetry breaking scale 
in the bulk is given by 
\begin{equation}
M_{sb} = \frac{1}{L}.
\end{equation}
If $L$ is $10\mu$m, we get the correct order of magnitude 
for the cosmological constant if $\rho_{\Lambda} \sim M_{sb}^4$.
In order to confirm this expectation, we should compute the 
effective potential for the radion which describes the size of 
the extra dimensions generated by supersymmetry 
breaking. The potential for the radion 
obtained by integrating out the bulk loops is given by
\begin{equation}
V(L) =\frac{c_2 M_6^2}{L^2} + \frac{c_3}{L^4}
(\log(M_6^2 L^2) +C).
\end{equation}
The calculation of $c_2$ depends on the details of the spectrum 
of the theory at $M_w$ and $c_2=0$ is critical for this model to 
work. If $c_2=0$, the potential leads to a natural 
realization of the quintessence model where the radion 
$L$ acts as a quintessence field. 

Thus SLED gives a consistent framework to address the cosmological 
constant problem and the dark energy model {\it provided
that} the self-tuning mechanism works and the supersymmetry 
breaking on the brane generates the desired potential for the radion,  
$V(L) \sim L^{-4}$. In SLED, the 6D Planck scale is 
supposed to be $M_w$ and the size of the extra dimensions
today are $L \sim 10 \mu$m. This leads to a lot of interesting 
phenomenology in local tests of gravity, collider physics
and so on \cite{Bur1, Bur2}. 
 
\section{Late-time acceleration in the braneworld}
A new twist to the cosmological constant problem is 
the late time acceleration of the Universe. 
The simplest way to realize this is to assume that 
a tiny amount of the cosmological constant is 
left after cancelling the vacuum energies. But the vacuum energy is 
typically more than 50 orders of magnitude larger than the 
observed value of the cosmological constant. 
Thus this is an incredible fine-tuning. 
Moreover, if the self-tuning idea works and the 
vacuum energy does not gravitate, it is 
in general difficult to realize the accelerated expansion 
of universe (see however the SLED proposal discussed in 
section III.C).
Alternatively, it is possible that there is 
no cosmological constant but that large-distance 
modification of GR accounts for the  
late-time acceleration. The braneword gravity 
provides a natural framework for the study of this 
possibility. For example in the model proposed by 
Dvali, Gabadadze and Porrtati (DGP), 4D GR
is modified on large scales \cite{Dva}. It is in fact possible 
to realize the accelerated expansion of the universe
without a cosmological constant \cite{Def, Def2}. This solution is known 
as the self-accelerating universe. We should note that 
in these attempts, we do not solve the old cosmological
constant problem. In addition, in the DGP model, 
the coincidence problem is not solved and we should 
introduce a fine-tuned dimensional parameter related 
to the scale of the cosmological constant,  
$\rho_{\Lambda} \sim 10^{-3}$ eV. However this is a novel 
alternative to dark energy models in GR and it 
gives a new perspective to approach the problem.

\subsection{Dvali-Gabadadze-Porrati model}
In the DGP model, gravity leaks off the
4D Minkowski brane into the 5D bulk
Minkowski spacetime at large scales. The 5D action describing the DGP model is given by
\be
S = \frac{1}{2 \kappa_5^2}\int d^5 x \sqrt{-g} R 
+ \frac{1}{2 \kappa_4^2} \int d^4 x \sqrt{-\gamma} \:\: 
{}^{(4)\!}R - \int d^4 x \sqrt{-\gamma} {\cal L}_m.
\ee
Instead of having the bulk cosmological constant and 
the tension on a brane as in the RS model,
there is an induced Einstein-Hilbert term 
on the brane.

On small scales, gravity is
effectively bound to the brane and 4D Newtonian 
dynamics is recovered to a good approximation. The transition
from 4D to 5D behaviour is governed by a crossover
scale
\begin{equation}
r_c = \frac{\kappa_5^2}{2 \kappa_4^2}.
\end{equation}
The weak-field gravitational potential behaves as
\begin{equation}
\Psi \sim \left\{ \begin{array}{lll} r^{-1} & \mbox{for} & r< r_c,
\\ r^{-2} & \mbox{for} & r> r_c. \end{array}\right.
\end{equation}
Unlike the RS model, gravity becomes 5D at large distances.
The DGP model was generalized by Deffayet to a
Friedman-Robertson-Walker brane in a Minkowski bulk \cite{Def2}.
The energy conservation equation remains the same as in
general relativity, but the Friedman equation is modified:
\begin{eqnarray}
&& \dot\rho+3H(\rho+p)=0\,,\label{ec} \\ && {H \over r_c}=
H^2 - {8\pi G_4 \over 3}\rho\,. \label{f}
\label{Fried}
\end{eqnarray}
The modified Friedmann equation shows that at late times in a 
CDM universe with $\rho \propto a^{-3} \to 0$, we have 
\begin{equation}
H\to H_\infty= {1\over r_c}\,.
\end{equation}
Since $H_0>H_\infty$, in order to achieve acceleration at late
times, we require $r_c\gtrsim H_0^{-1}$, and this is confirmed by
fitting SN observations \cite{Def4}.
Like the LCDM model, the DGP model has simple background 
dynamics, with a single
parameter $r_c$ to control the late-time acceleration.

\begin{figure}[httb]
\centerline{
\includegraphics[width=9cm]{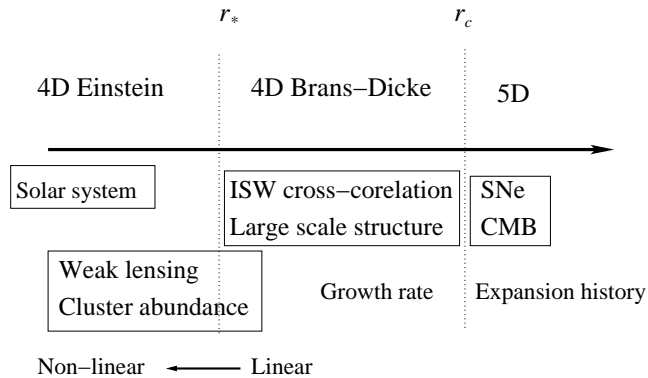}}
\caption{Summary of the behaviour of gravity in 
the DGP model. At large scales $r>r_c$, the theory 
is 5D. On small scales $r<r_c$, gravity becomes 4D but 
the linearized theory is described by a Brans-Dicke 
theory. This affects the large scale structure (LSS) and 
the Integrated Sachs-Wolfe (ISW) effect and its 
cross-correlation to LSS. Below the Vainstein radius
$r < r_*$, the theory approaches GR. This transition
can be probed by weak lensing and cluster abundance 
as the non-linear dynamics is important for these measures. 
The solar system tests also provide constraints on the model
in the 4D Einstein phase. 
}
\label{fig4}
\end{figure}

On small scales, the Newtonian potential behaves as 4D. The 
Friedmann equation also shows that the universe behaves as 4D 
at early times, $Hr_c \gg 1$. However, the recovery of GR is 
very subtle in this model \cite{Def3}. In fact, 
although the weak-field gravitational potential behaves as 4D on 
scales smaller than $r_c$, the linearized gravity is {\it not} described 
by GR. This is because there is no normalized 
zero-mode in this model and 4D gravity is recovered as 
a resonance of the massive KK gravitons. The massive graviton 
contains 5 degrees of freedom compared with 2 degrees of freedom 
in a massless graviton. One of them is a helicity-0 polarization. 
Due to this scalar degree of freedom, linearized gravity is described 
by Brans-Dicke (BD) gravity with vanishing BD parameter in the 
case of Minkowski spacetime. Thus this model would be excluded by 
solar system experiments. However, the non-linear interactions 
of the scalar mode becomes important on larger scales than expected \cite{Def3, 
Lue, Gru, Tan}. 
Let us consider a static source with mass $M$. Gravity becomes non-linear 
near the Schwarzshild radius $r_g= 2 G M$. However, the scalar mode becomes 
non-linear at $r_* = (r_g r_c^2)^{1/3}$ (the Vainstein radius) 
which is much larger than $r_g$ if $r_c \sim H_0^{-1}$. In fact, for the Sun $r_*$ is much 
larger than the size of the solar system. 
A remarkable finding is at once the scalar mode becomes non-linear,
GR is recovered. This non-linear shielding of the 
scalar mode is crucial to escape from the tight solar system constraints.  
Fig.~\ref{fig4} summarizes the behaviour of gravity in the DGP model
(see \cite{LueR} for a review on the DGP model).

\subsection{Observational constraints on the self-accelerating universe}
The self-accelerating universe provides useful example where 
we can study how various observations can be combined to test the model. 
It also provides a possibility to find a failure of 
GR at cosmological scales. A key is the complicated behaviour of 
gravity. We have various cosmological observations that 
cover various scales. Then combining the various data sets,
we can probe the complicated behaviour of gravity in this 
model. A central question is whether we can distinguish the DGP 
model from dark energy models in GR.

\subsubsection{Expansion history}
The first question is whether one can distinguish between the 
self-accelerating universe and the simple $\Lambda$CDM model in GR.
Both models have the same number of parameters and phenomenologically
both theories have the same simplicity. In terms of density 
parameters, the Friedmann equation 
in the $\Lambda$CDM model is given by
\begin{equation}
\Omega_M + \Omega_{\Lambda} + \Omega_K =1.
\end{equation}
On the other hand in the DGP, we have \cite{Def4}
\begin{equation}
\Omega_M + 2 \sqrt{\Omega_{r_c}} \sqrt{1-\Omega_K} + \Omega_K
=1,
\end{equation}
where we defined 
\begin{equation}
\Omega_{r_c} = \frac{1}{4 H_0^2 r_c^2}.
\end{equation}

\begin{figure}[h]
\begin{center}
\includegraphics[width=8cm]{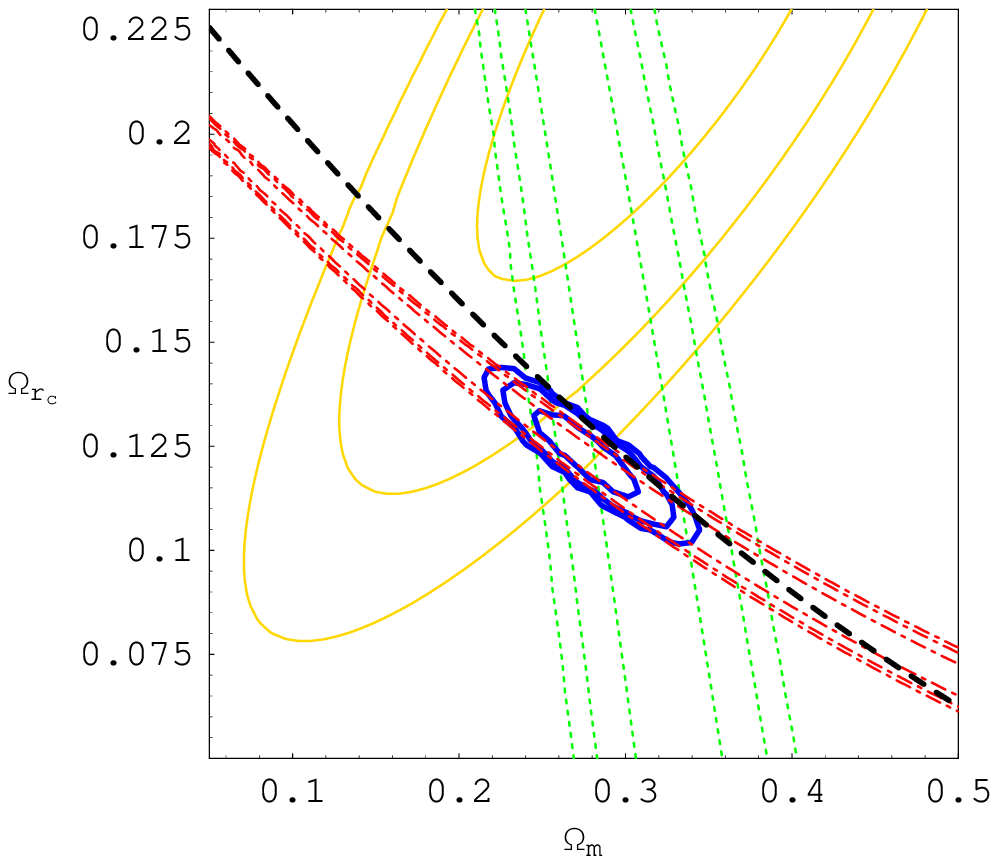}\quad
\includegraphics[width=8cm]{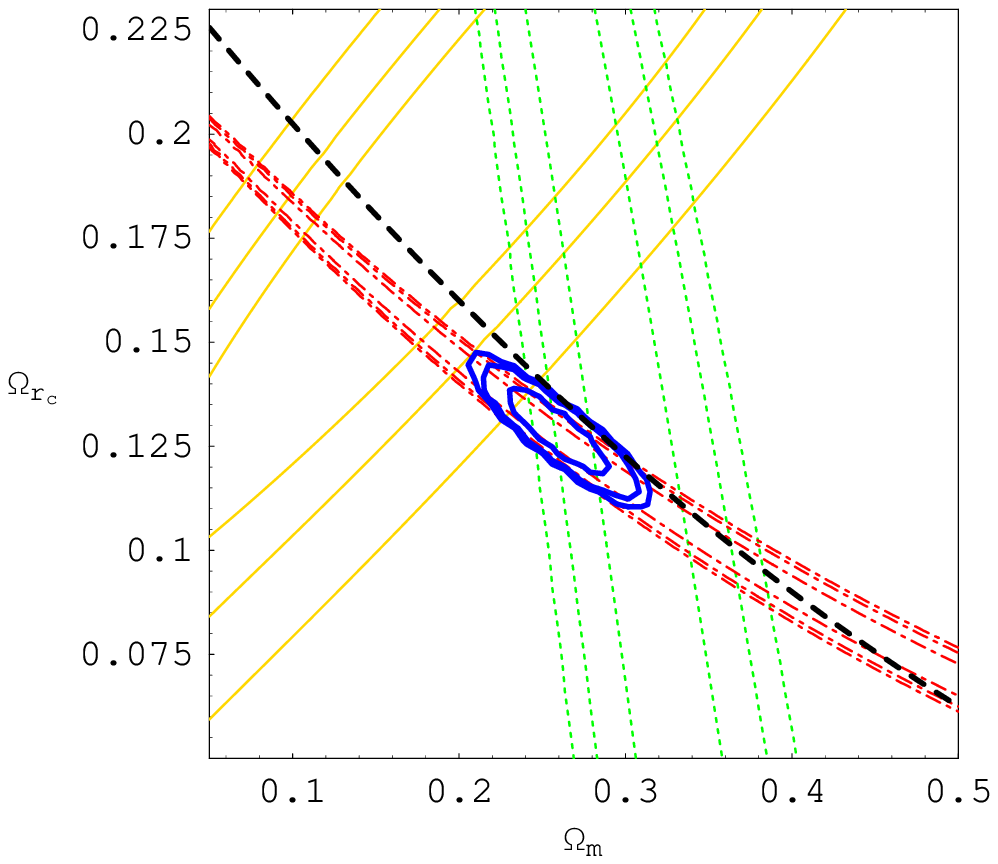}
\caption{Joint constraints [solid thick] on DGP models from
the SNe data [solid thin], the BO measure $A$ [dotted] 
and the CMB shift parameter $S$ [dot-dashed]. The
left plot uses SNe Gold data, the right plot uses SNLS data. The
thick dashed line represents the flat models,
$\Omega_K=0$. From \cite{Mar2}.}\label{sac}
\end{center}
\end{figure}

\begin{figure}[h]
  \begin{center}
  \epsfysize=3.3truein
  \epsfxsize=3.3truein
    \epsffile{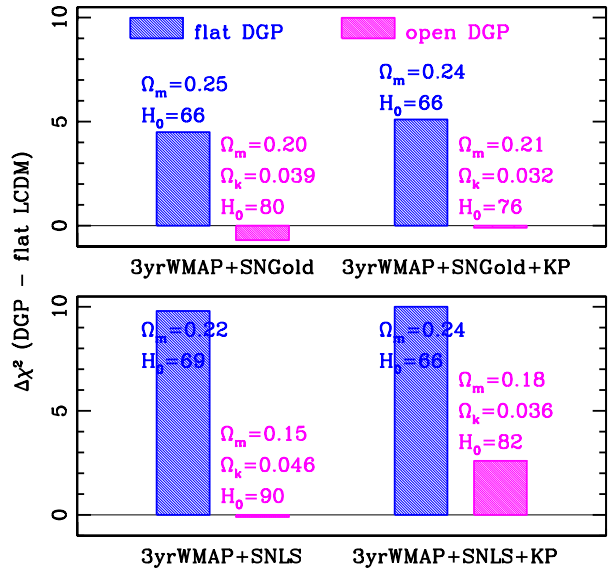}
    \caption{
The $\Delta\chi^2$ between the best fit
flat and open DGP versus that of a flat $\Lambda$CDM model.
The Gold supernova (SN) data set is used
in the top panel and the SNLS SN data set is used in the bottom panel.  The
DGP model requires curvature and a high Hubble constant.  With the
addition of Key Project (KP) direct Hubble constant measurements, open
DGP is a marginally poorer fit to the data than flat $\Lambda$CDM.
From \cite{Son}}
\label{fig:dchi}
\end{center}
\end{figure}

In order to constrain the density parameters, we can 
combine data from supernovae, the cosmic microwave 
background shift parameter, and possibly the baryon
oscillation peak \cite{Fai, Mar2, Son, Barg, Ryd, Laz}. 
Interestingly, the current observations already give 
us a hint how we can distinguish the models. While 
the $\Lambda$CDM model fits the three data sets 
comfortably, there is some tension between the data 
and DGP (Fig.~\ref{sac})\cite{Mar2}. It is suggested that a slightly open 
universe can fit the data set better in the DGP (Fig.~\ref{fig:dchi}) 
\cite{Son}.

%%%%%%%%%%%%%%%%%%%%%%%%%%%%%%

\begin{figure}[h]
\begin{tabular}{cc}
\includegraphics[height=6cm,width=8cm]{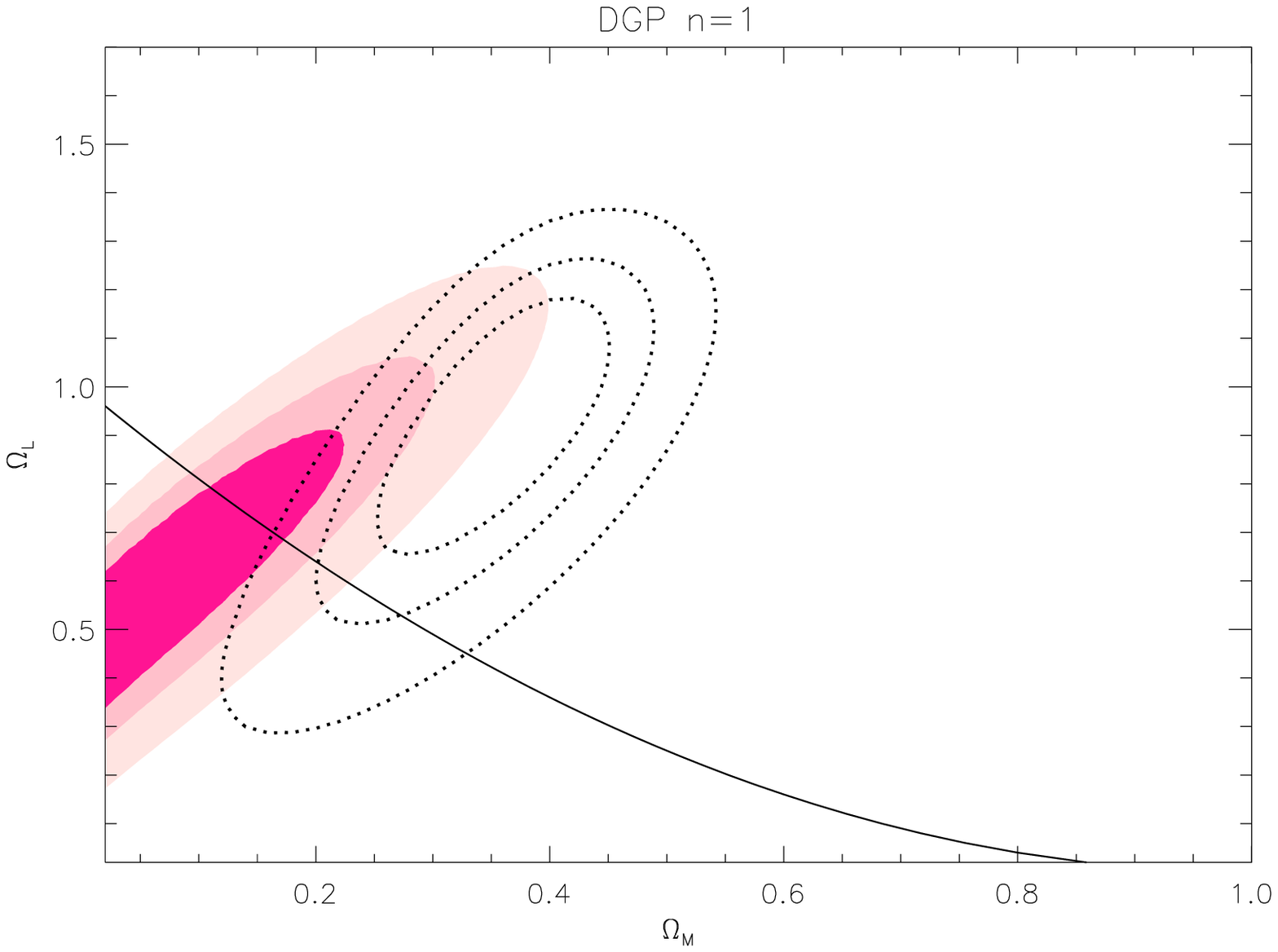}& 
\includegraphics[height=6cm,width=8cm]{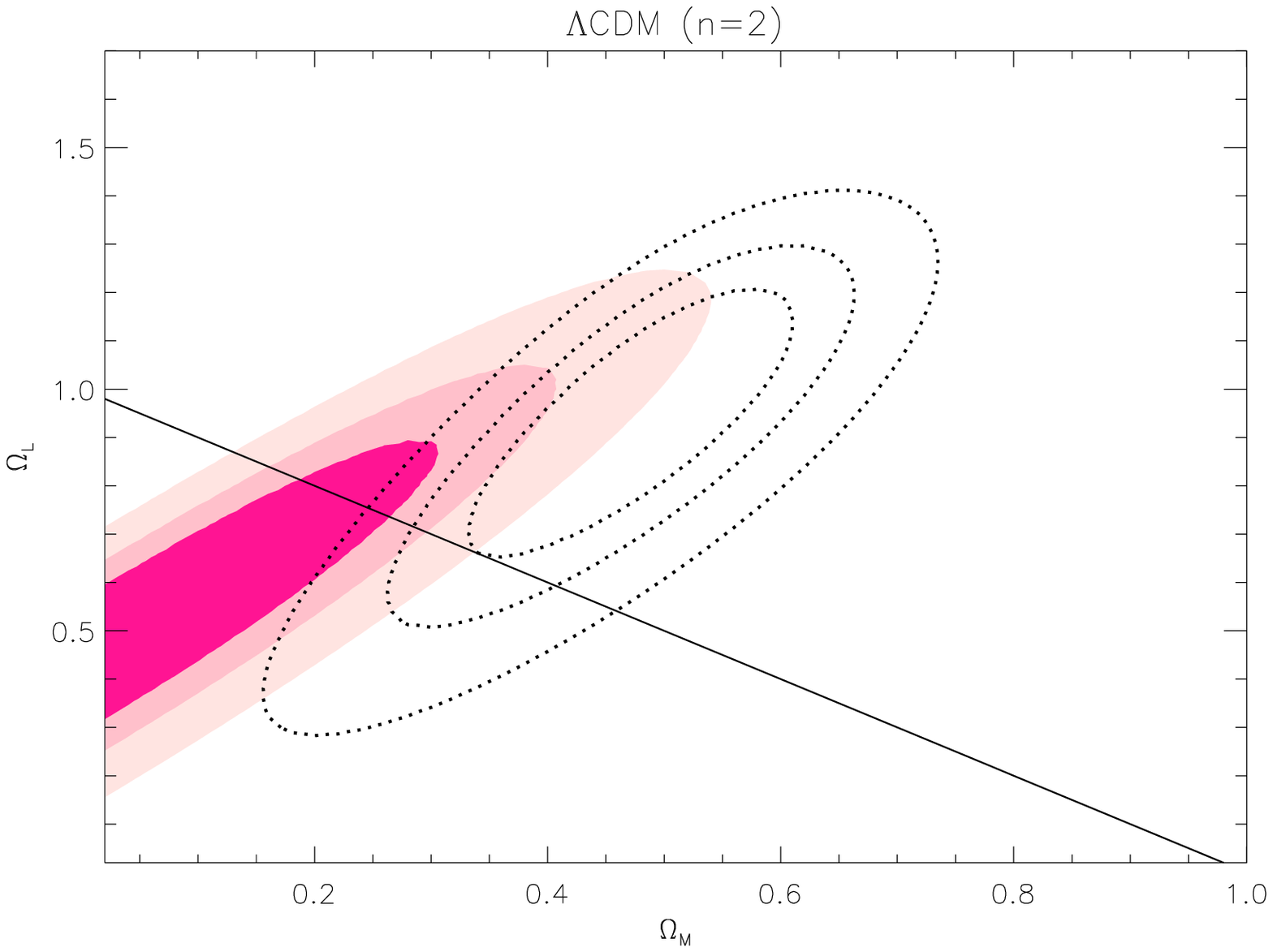}\cr
\end{tabular}    
\caption{Comparison between the results of fitting DGP and $\Lambda$CDM to the SNLS and ESSENCE supernova data set (filled in 68\%, 95\% and 99\% confidence regions) and the Riess 07 Gold set (dotted lines). The solid black line corresponds to spatially flat universes. From \cite{Ryd}.
\label{Riess} }
\end{figure}

Note that the baryon acoustic oscillation measure requires 
the knowledge of the power spectrum thus the knowledge 
of perturbations. Precisely speaking, the analysis 
must be redone for the DGP model. We expect that 
only small corrections are involved, but this problem 
must be addressed. 
The conclusion also seems to depend on the data set for 
supernovae (Table~I and II) \cite{Mar2}. This is also true using the latest results 
from the ESSENCE and SNLS supernova data set and 
the Riess 07 Gold set (Fig.~{\ref{Riess}}) \cite{Ryd}. 
%%%%%%%%%%%%%%%%%%%%%%%%

\begin{center}
\begin{table}[h]
\begin{tabular}{|c|c|c|c|c|} \hline
 & best-fit  & best-fit & best-fit &$\chi^2$
 \\ & acceleration & density & curvature& value \\
 & parameter & parameter &parameter &  \\\hline
 DGP & $\Omega_{rc}=0.125$ & $\Omega_m=0.270$ & $\Omega_K=+0.0278$ & 185.0\\
 ~LCDM~ & $~\Omega_\Lambda=0.730~$ &
 $~\Omega_m=0.285~$ & $~\Omega_K=-0.0150~$& $~177.8~$\\ \hline
\end{tabular}
\caption{Best-fit parameters from SNe-CMB shift-Baryon Oscillation
constraints, and $\chi^2$ values, for the DGP and LCDM models.
The Gold data is used for the SNe. From \cite{Mar2}}
\end{table}
\end{center}
\begin{center}
\begin{table}[h]
\begin{tabular}{|c|c|c|c|c|} \hline
 & best-fit  & best-fit & best-fit &$\chi^2$
 \\ & acceleration & density & curvature& value \\
 & parameter & parameter &parameter &  \\\hline
 DGP & $~\Omega_{rc}=0.130~$ & $~\Omega_m=0.255~$ & $~\Omega_K=+0.0300~$ & $~128.8~$\\
 ~LCDM~ & $~\Omega_\Lambda=0.740~$ &
 $~\Omega_m=0.270~$ & $~\Omega_K=-0.0100~$& 113.6\\ \hline
\end{tabular}
\caption{As in Table~I for the Legacy SNe data. From 
\cite{Mar2}}
\end{table}
\end{center}

\begin{figure}[h]
\begin{center}
\includegraphics[width=8cm,angle=0]{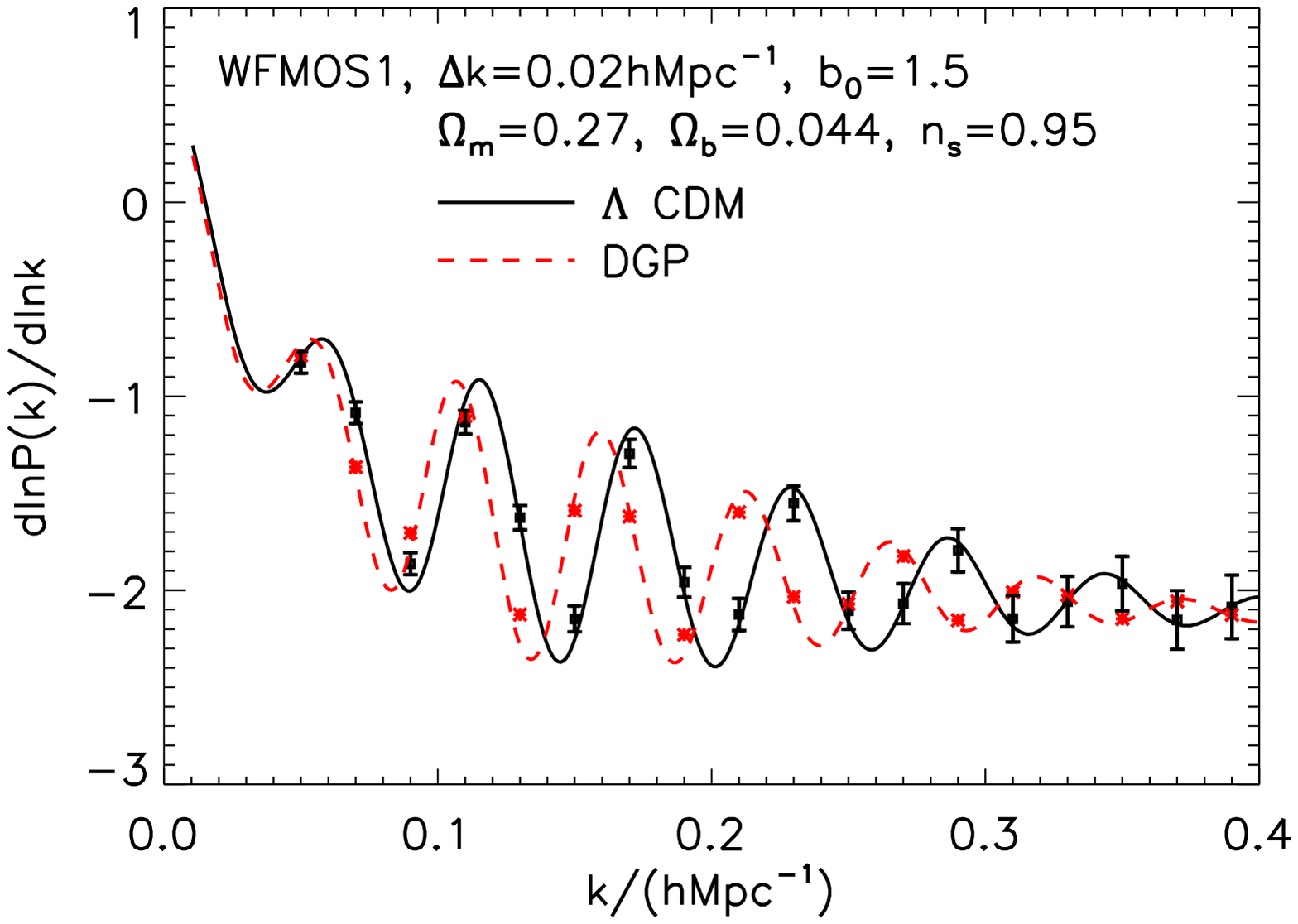}
\caption{Theoretical predictions for ${d \ln{P(k)}/d \ln k}$ 
assuming a sample WFMOS, where $P(k)$ is the power spectrum 
of galaxies. 
The squares with error bars are 
evaluated with the simple simulation of the power spectrum
for the $\Lambda$CDM model. The asterisks are the DGP model,
but the error bars, which are almost the same  
as that of the $\Lambda$CDM model, are omitted for simplicity. 
Theoretical curves are the DGP model (dashed red curve) and the 
$\Lambda$CDM model (solid black curve). 
The parameters are $n_s=0.95$, 
$\Omega_b=0.044$, $\Omega_m=0.27$ and the linear bias is taken as 
$b_0=1.5$. From \cite{Yam}}
\label{fig10}
\end{center}
\end{figure}

In the future, precision data will enable us to 
distinguish between the DGP and the $\Lambda$CDM
more clearly. Fig.~\ref{fig10} shows the prediction of the baryon 
acoustic peak oscillation observed by a future WFMOS survey 
which is assumed to contain $2.1 \times 10^6$ galaxies, over 
$200$ deg$^2$, at $0.5 <z<1.3$ \cite{Yam}. Clearly 
the difference between the two is much larger than the 
error bars. 

%%%%%%%%%%%%%%%%%%%%%%%%%%%%%%%%%%%%%%%%%%%%%%%%%%%%%%%%%%%
%%%%%%%%%%%%%%%%%%%%%%%%%%%%%%%%%%%%%%%%%%%%%%%%%%%%%%%%%%%

\subsubsection{Linear growth of structure}
Although the DGP model can be distinguished from the 
$\Lambda$CDM model, background tests will never 
distinguish the DGP model from dark energy models 
in GR. This is because there always exists a dark energy model 
in GR that has exactly the same expansion history as in DGP. 
In fact as far as the background evolution of the Universe is
concerned, the DGP is equivalent to the dark energy model whose 
equation of state is given by \cite{Lue2}
\begin{equation}
w = - \frac{1}{1 +\Omega_m(a)}.
\end{equation}
For small red-shift, this is well fitted by 
$w = w_0 + w_a (1-a)$ where $w_0=-0.78$ and $w_a =0.32$ if 
$\Omega_m=0.3$ today \cite{Lin}. 
Then we cannot distinguish the DGP from the dark energy model
in GR.

However, even if the background dynamics is the same, this 
does not mean that the dynamics of perturbations is the same. 
Koyama and Maartens obtained the solutions for metric perturbations 
on sub-horizon scales by consistently solving the 5D perturbations under 
quasi-static approximations \cite{Koy1}.
Scalar metric perturbations are given in longitudinal
gauge by
\begin{equation}\label{smp}
ds^2= -(1+2 \Psi)dt^2 + a^2 (1+2 \Phi) \delta_{ij} dx^i dx^j \,,
\end{equation}
and the perturbed energy-momentum tensor for matter is given by 
\begin{equation}
\delta T^{\mu}_{\nu} = 
\left(
\begin{array}{cc}
 -\delta \rho & a \delta q_{,i} \\
 -a^{-1} \delta q^{,i}  & \delta p \: 
 \delta^{i}_{\:\: j} \\
\end{array}
\right).
\end{equation}
The solutions for the brane metric perturbations are \cite{Koy1}
\begin{eqnarray}
\frac{k^2}{a^2} \Phi &=& 4\pi G_4 \left(1- \frac{1}{3 \beta} \right)
\rho \delta, \label{solphi}\\
\frac{k^2}{a^2} \Psi &=& -4\pi G_4 \left(1 + \frac{1}{3 \beta}
\right) \rho \delta, \label{solpsi}
\end{eqnarray}
where
\begin{equation}
\beta =1 -2 r_c H \left(1+ \frac{\dot{H}}{3 H^2} \right),
\end{equation}
and 
\begin{equation}
\delta = \delta \rho -3 H \delta q.
\end{equation}
This agrees with the results obtained by Lue, Scoccimarro and Starkman. 
They find spherically symmetric solutions 
by closing the 4D equations using an anzatz for the metric and checking
in retrospect that the obtained solutions satisfy regularity in the 
bulk. It was shown that the solutions (\ref{solphi}) and (\ref{solpsi}) 
are uniquely determined by the regularity condition 
in the bulk within our approximations. 

The modified Poisson equation~(\ref{solphi}) shows the suppression
of growth. The rate of growth is determined by $\delta$, and for CDM,
\begin{equation}
\ddot{\delta} + 2 H \dot{\delta}=-\frac{k^2}{a^2}
\Psi\,,
\end{equation}
which leads to
\begin{equation}\label{dpe}
\ddot{\delta} + 2 H \dot{\delta}=4\pi G_4 \left(1 +
\frac{1}{3 \beta} \right) \rho \delta\,.
\end{equation}
Thus the growth rate receives an additional modification from the
time variation of Newton's constant through $\beta$. 

\begin{figure}[h]
\centerline{
\includegraphics[width=8cm]{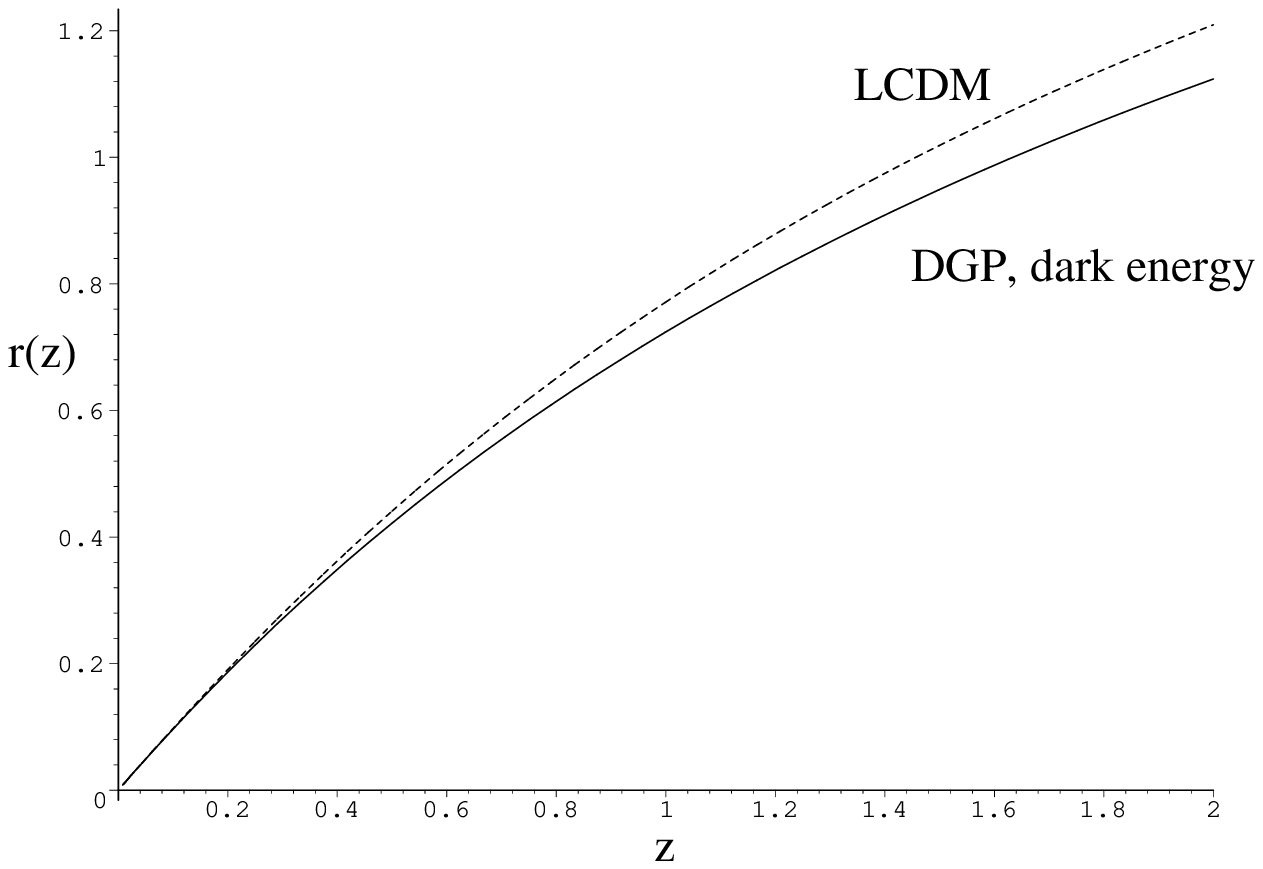} \quad
\includegraphics[width=8cm]{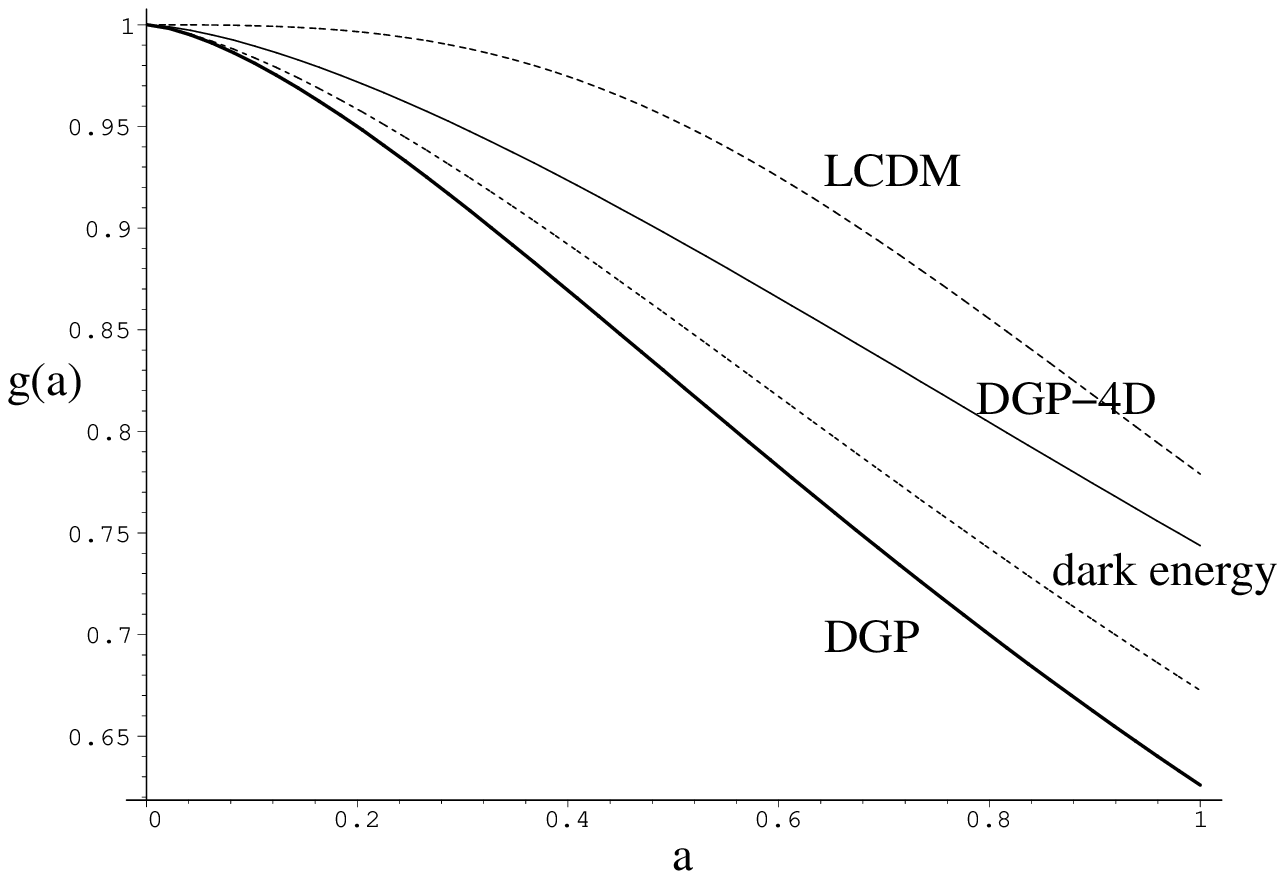}}
\caption{The comoving distance r(z) is shonw for 
$\Lambda$CDM (long dashed), DGP (solid, thick) and 
the equivalent GR dark energy model on the left. 
On the right, the growth history $g(a)=\delta(a)/a$ is shown for 
LCDM (long dashed) and DGP (solid, thick). The growth history for a 
dark energy model (short dashed) is also shown, with the same 
expansion history as DGP. Due to the time variation of Newton's 
constant through $\beta$ in Eq.~(\ref{dpe}), the growth factor $g(a)$ receives 
an additional suppression compared with the dark energy model. 
DGP-4D (solid, thin) shows the incorrect 
result in which the inconsistent assumption 
is adopted. We set the density parameter for matter today as 
$\Omega_{m0}=0.3$. From \cite{Koy1}}
\label{KM}
\end{figure}

In Fig.~\ref{KM}, we show the linear growth factor $\delta/a$ for the DGP
model, and compare it with $\Lambda$CDM and with the GR 
dark energy model whose background evolution matches 
that of the DGP model. We also show the incorrect DGP result, 
in which the inconsistent assumption of neglecting 5D perturbations
is effectively adopted \cite{Son0}.
This inconsistent assumption has been made in various treatments
but it leads to unreliable results. 
The correct equations for subhorizon density perturbations are
crucial for meaningful tests of DGP predictions against structure
formation observations. This highlights the fact that the growth rate
is very sensitive to the modification of gravity.

There are several observations that can probe the 
growth of structure. Weak lensing measures the deflection 
of light generated by matter fluctuations 
(see \cite{Mun} for a review). The deflection 
potential is given by
\begin{equation}
\phi = \Phi + \Psi.
\end{equation} 
We can relate $\phi$ to the matter overdensity $\delta$:
\begin{equation}
\phi = \frac{8 \pi G_4 a^2}{k^2} \delta.
\end{equation}
Interestingly, this formula in the DGP is the same as the one in GR.
However, the change of the growth rate leads to a different
prediction of weak lensing. We should note that current 
observations measure weak lensing sourced by matter fluctuations 
in the non-linear regime. The solutions (\ref{solphi}) can be
applied only to linear perturbations and there is no 
justification to use the linear growth rate and predict the non-linear 
power spectrum using the mapping formula developed in GR. 
We will come back to this issue in section IV.B. 

Another probe is the Integrated Sachs-Wolfe (ISW)
effect. This is determined by the time variation of the 
deflection potential $\dot{\phi}$. On large scales, 
we should deal with the truly 5D effects and the quasi-static
solutions are not applicable. There is some progress 
to deal with fully dynamical perturbations by adopting 
a scaling ansatz to solve the 5D equations \cite{Saw}. 
They find that the quasi-static solution is an attractor on subhorizon 
scales. The ISW effects are sub-dominant compared with 
the primordial anisotropies formed at the last scattering 
surface. In order to extract the ISW effects, it is proposed 
to take cross correlation between the matter distributions 
and the CMB \cite{Cri1, Bou, Pog}. 
It was shown that the quasi-static solution is valid to 
calculate the corss correlation for large 
$\ell$ where a signal is maximized \cite{Saw}.
The growth function $g(a)$ changes at earlier times in 
the self-accelerating universe than in the $\Lambda$CDM 
model. This gives a larger signal in the cross correlation
at high redshift. Thus higher red-shift galaxies can test the 
predictions in the self-accelerating universe with 
high significance. 

\begin{figure}[t]
\centerline{
\epsfxsize=5.3truein\epsffile{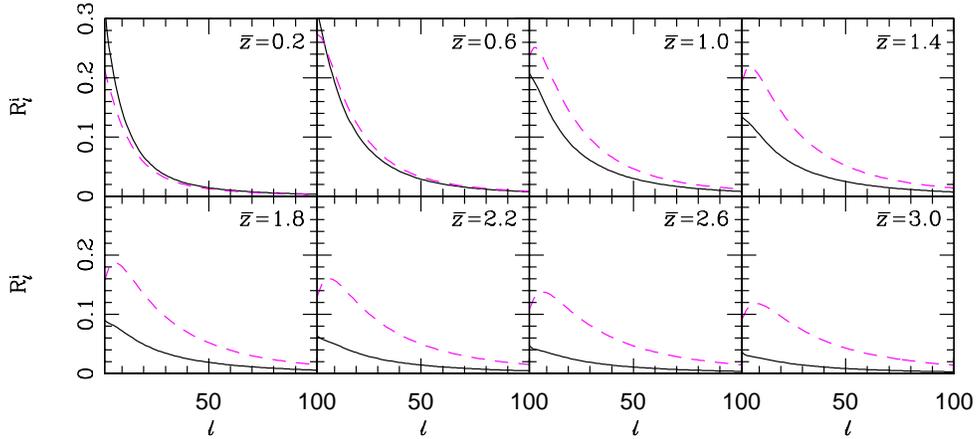}$\,$}
\caption{
The galaxy-ISW cross-correlation coefficient
$R^i_{\sc l}$ in each galaxy bin from $z=0$ to $z=3$.
Solid curves denote flat $\Lambda$CDM
and dashed curves denote open DGP.  Note the much larger
correlation at high $z$ in open DGP. From \cite{Saw}.
}
\label{fig:R}
\end{figure}

\begin{figure}[t]
\begin{center}
\includegraphics[width=2.5in,height=3.2in,angle=-90]{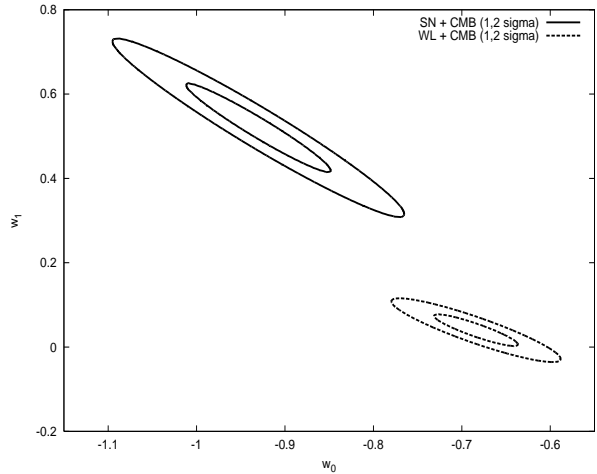}
\caption{\label{fig:eossnwl_a} 
    Equations of state found using two
    different combinations of data sets.  Solid contours are
    for fits to SN Ia and CMB data, while dashed contours are
    for fits to weak lensing and CMB data.  The significant
    difference (inconsistency) between the equations of state found using
    these two combinations is a signature of the DGP
    model. The inconsistency is an observational detection
    of the underlying modified gravity DGP model 
	(assumed here to generate the data). From \cite{Ish}.
}
\label{ish}
\end{center}
\end{figure}

Hence, structure formation tests are essential for breaking
the degeneracy with dark energy models in GR \cite{Lue3, Ish, Koy2, 
Zha, Lin2}.
The distance-based SN observations draw only upon
the background 4D Friedman equation~(\ref{f}) in DGP models, and
therefore there are quintessence models in GR that
can produce precisely the same SN redshifts as DGP. By
contrast, structure formation observations require the 5D
perturbations in DGP, and one cannot find equivalent GR models.
This leads to an exciting possibility to find a failure of GR \cite{Ish}.
Suppose that our Universe is described by the DGP model. However, 
astronomers still try to fit the data by dark energy models in 
GR. For example, they use the parametrization of the equation of 
state of dark energy 
\begin{equation}
w = w_0 + w_1 z.
\end{equation}
Combining SN observations, CMB shift parameter and weak lensing,
there appears an inconsistency. This is because weak lensing 
probes the growth of structure and the growth rate in the 
DGP model cannot be 
fitted by the growth rate in GR models given the same expansion 
history. Fig.~\ref{ish} demonstrates this possibility. 

\begin{figure}[h]
\centerline{
\includegraphics[clip,scale=0.6]{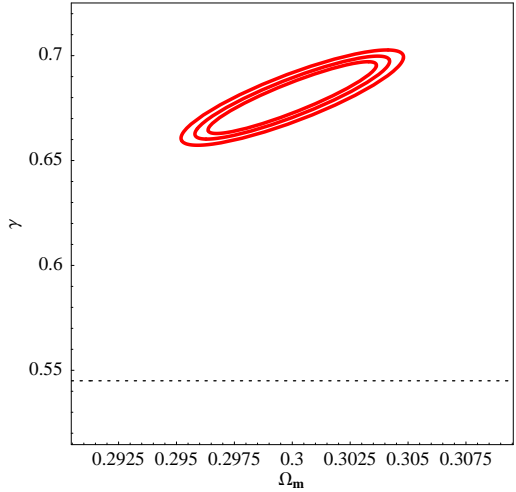}}
\caption{\label{fig:dgp}Confidence regions at 68\% for the benchmark survey
$z_{mean}=0.9,d=35$ (outer contour) and for $d=50,75$ (inner contours)
for DGP. The dotted line represents the $\Lambda$CDM value. From \cite{Ame}.}
\label{gamma}
\end{figure}

In order to quantify the difference in the growth rate,  
it is convenient to parametrize the growth rate as \cite{Lin}
\begin{equation}
g(a) = \exp \left\{\int^a_0 d \ln a (\Omega(a)^{\gamma} -1 )\right\}.
\end{equation}
In a quintessence model, $\gamma$ is well approximated by 
\begin{equation}
\gamma(w) = 0.55 + 0.05 (1 + w(z=1)).
\end{equation}
In the DGP, $\gamma$ is well approximated as $\gamma = 0.68$ \cite{Lin}.
Recently, several authors tried to estimate how accurately we 
can constrain $\gamma$ using weak lensing in future surveys
\cite{Hea, Yam2, Ame}. 
These results suggest that in the future, we will be able to discriminate  
$\Lambda$CDM and the DGP model from the difference in the growth rate. 
Fig.~\ref{gamma} shows the constraint on $\gamma$ for the 
DGP model assuming 
the `bench mark' survey on weak lensing, where 
the mean redshift is $z_{\mbox{\small{mean}}} =0.9$
and the number of sources per arcmin$^2$ is $d=35, 50, 75$ \cite{Ame}. 

However, as we mentioned before, the weak lensing measure 
requires knowledge of the non-linear power-spectrum. In the DGP, this is 
a subtle problem. The DGP approaches GR on small scales. 
This is essential to evade the tight constraints from 
the solar system experiments. 
The non-linear power spectrum would be sensitive to this 
transition from Brans-Dicke linear theory to GR non-linear 
theory. The analyses so far have used the simple mapping 
formula developed in GR to derive the non-linear power spectrum. This approach 
could be inconsistent. Nevertheless, the conclusion 
that we will be able to distinguish the difference in the 
growth of structure would be valid and this is a very exciting 
possibility that we can achieve in future observations. 

\subsubsection{Non-linear structure formation}
For quasi-static perturbations, it is possible to extend the 
linear result to non-linear perturbations by taking into account 
partially the non-linear effects of gravity. A key is the so called brane 
bending mode \cite{Tan, Lut, Nic}. This mode describes the perturbations of the location
of the brane and mediates an additional scalar interaction. 
In the linear regime, this is the scalar mode that makes the 
theory of BD type. This scalar mode becomes non-linear 
on much larger scales than gravity. In terms of the 
brane bending mode $\varphi$, the effective equations
on the brane are given by \cite{Koy3}
\begin{eqnarray}
\frac{2}{a^2} \nabla^2 \Phi &=& -8 \pi G_4 \delta \rho 
+ \frac{1}{a^2} \nabla^2 \varphi, 
\label{einstein:tt} \\
\Psi +\Phi &=& \varphi,
\label{einstein:traceless} 
\end{eqnarray}
where the equation of motion for $\varphi$ is given by
\begin{equation}
3 \beta(t)
\frac{\nabla^2}{a^2} \varphi 
+ \frac{r_c^2}{a^4} 
\left[      
(\nabla^2 \varphi)^2 - (\nabla_i \nabla_j \varphi)^2 
\right]=8 \pi G_4 \delta \rho.
\label{eq:phi}
\end{equation}
Again these equations are derived by properly solving 
the 5D equations and imposing the regularity condition 
in the bulk. Here we assume gravity is linear, $\Psi, \Phi \ll 1$,
but we take into account the second order effects of $\varphi$.
Note that the coefficient of the second order terms is 
given by $r_c^2$. As we take 
$r_c \sim H_0^{-1}$, the second order terms can be 
comparable to the linear term even if gravity remains 
linear. 

These non-linear equations are difficult to solve in general. 
If we assume spherical symmetry, the solution for  
$\varphi$ is given by
\begin{equation}
\frac{d\varphi}{dr} = \frac{r_g}{r^2} \Delta(r), 
\quad \Delta(r) = \frac{2}{3 \beta} \left( \frac{r}{r_*} \right)^3 
\left(\sqrt{1+ \left(\frac{r_*}{r}\right)^3} -1 \right),
\label{phir}
\end{equation}
where
\begin{equation}
r_*= \left(\frac{8 r_c^2 r_g}{9 \beta^2} \right)^{1/3},
\end{equation}
and $r_g$ is the Schwarzschild radius $r_g = 2 G_4 M$. 
The solutions for $\Phi$ and $\Psi$ are obtained as 
\begin{eqnarray}
\Phi &=& \frac{r_g}{2 r} + \frac{\varphi}{2}, \\ 
\Psi &=& -\frac{r_g}{2 r} + \frac{\varphi}{2}.
\end{eqnarray}
For $r>r_*$, we recover the 
solutions for linear perturbations (\ref{solphi}).
For $r < r_*$, the solutions for metric perturbations 
are given by \cite{Lue2}
\begin{eqnarray}
\Phi &=& \frac{r_g}{2r} + \frac{1}{\beta} 
\sqrt{\frac{\beta^2 r_g r}{2r_c^2}}, \\
\Psi &=& -\frac{r_g}{2r} + \frac{1}{\beta} 
\sqrt{\frac{\beta^2 r_g r}{2 r_c^2}}.
\label{einsteinphase}
\end{eqnarray}
In this region, the corrections to the solution in 4D GR 
are suppressed, so that Einstein gravity 
is recovered. The radius $r_*$ is the Vainstein radius
in the cosmological background. 

\begin{figure}[h]
\centerline{\epsfxsize=8cm\epsffile{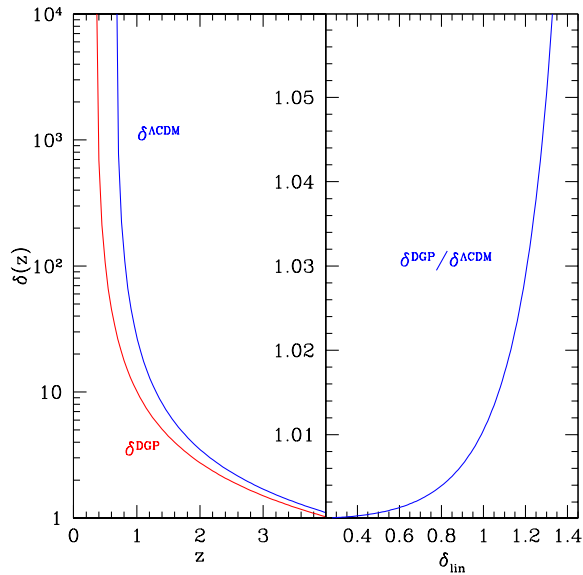}}
\caption{Numerical solution of spherical collapse. 
The left panel shows the evolution for a spherical perturbation with $\delta_i=3\times 10^{-3}$ 
at $z_i=1000$ for $\Omega_m=0.3$ in DGP gravity and in $\Lambda$CDM. 
The right panel shows the ratio of the solutions once they are both expressed 
as a function of their linear density contrasts. From \cite{Lue2}}
\label{fig:SphColl}.
\end{figure}

The conservation of the energy momentum tensor holds as in 
GR. Then the continuity equation and the Euler equation are the same 
as in GR:
\begin{eqnarray}
\frac{\partial \delta}{\partial t} + \frac{1}{a} \nabla^i (1+\delta) 
v_i &=&0, 
\label{matter1}\\
\frac{\partial v_i}{ \partial t} + \frac{1}{a}
(v^j \nabla_j) v_i  + H v_i 
&=& - \frac{1}{a} \nabla_i \Psi,
\label{matter2}
\end{eqnarray}
where $v_i$ is the velocity perturbation of dark matter.
Eqs.~(\ref{einstein:tt}), (\ref{einstein:traceless}), (\ref{eq:phi})
(\ref{matter1}) and (\ref{matter2}) form a closed set of equations
that has to be solved to address the non-linear structure 
formation problem in the DGP model. 
In order to see how GR is recovered dynamically, 
let us consider the evolution of a spherical top-hat perturbation $\delta(t,r)$ 
of top-hat radius $R_t$,
where $\rho(t,r)=\rho(t)(1+\delta)$ is the full density distribution 
and $\rho(t)$ is the background density \cite{Lue2}.
The Newtonian potential 
$\Psi$ dominates the geodesic evolution of overdensity. 
Then the evolution equation for the over-density $\delta$
is given by
\begin{equation}
\ddot{\delta} - \frac{4}{3} \frac{\dot{\delta}^2}{1+\delta}+2H\dot{\delta} = 4\pi G \rho \delta (1+\delta) \left[ 1 + \frac{2}{3\beta} \frac{1}{\epsilon} \left( \sqrt{1+\epsilon}-1\right)\right]\ ,
\label{sphc}
\end{equation}
\begin{equation}
\epsilon\equiv \frac{8r_c^2 r_g}{9\beta^2R_t^3} =
\frac{8}{9} \frac{(1+\Omega_m)^2}{(1+\Omega_m^2)^2} \Omega_m \delta.
\end{equation}
In the linear regime, $\delta \ll 1$, $\epsilon \ll 1$, 
we recover the linear evolution of the overdensity, Eq.~(\ref{dpe}). 
On the other hand for $\epsilon \gg 1$, the right 
hand side of Eq.~(\ref{sphc}) becomes the same as 
in GR and the dynamics of the non-linear collapse 
becomes the same as in GR. Fig.~\ref{fig:SphColl}
shows the behaviour of $\delta$ in the DGP 
compared with the $\Lambda$CDM model. 

For the non-spherically symmetric case, we need to solve the equations 
numerically. We should emphasize again that the analysis of the 
non-linear transition of the theory to GR is essential for 
the prediction of weak lensing and this is an outstanding 
open problem.

\subsection{Theoretical consistency of the DGP model}
Although the DGP model offers a concrete example for a 
modified gravity alternative to dark energy, this model 
is not free from problems. In fact, this model demonstrates 
the difficulties of modifying GR at large distances. 
One of the problems is related to the non-linearity of the 
scalar mode. The non-linear interactions of the scalar 
mode become important at the Vainstein length $r_*
\sim (r_c^2 r_g)^{1/3}$. If we consider a Planck 
scale mass particle, this length is given by 
$\Lambda_c^{-1} = (r_c^2 M_{pl}^{-1})^{-1/3}$, which is 
$\sim 1000$ km for $r_c \sim H_0^{-1}$. This  
defines the length below which quantum corrections for 
the scalar mode become important. Thus $\Lambda_c$
plays the same role as the Planck scale in GR.
Then below the length $\Lambda_c^{-1}$, 
the classical theory loses its predictability. 
This is known as the strong coupling problem \cite{Rub2, Lut, Nic}.
There have been debates whether this 
is indeed a problem or not \cite{Dva2}. It is suggested that there exists 
a consistent choice of counter-terms for which the 
model remains calculable \cite{Nic}. 

The most serious problem in this model 
is that there are ghost-like excitations around the self-accelerating 
universe \cite{Lut, Nic, Koy4, Gor, Cha, Izu}. 
In fact the growth rate already manifests this problem. The solution
for the linearized perturbations is described by a BD 
theory with BD parameter given by \cite{Lue, Koy1}
\begin{equation}
\omega = \frac{3}{2} (\beta-1).
\end{equation}
For large $Hr_c$, $\beta$ is always negative. 
In fact, if $\omega < -3/2$, the BD scalar field has 
the wrong sign for its kinetic term and it becomes 
a ghost. For de Sitter spacetime, the condition 
$\omega < -3/2$ implies $Hr_c >1/2$. This is exactly
the condition that there exists a ghost in the theory. 
We can understand the extra suppression of the 
growth rate as due to the repulsive force mediated 
by the ghost. If we avoid the negative norm state when 
quantizing the theory with ghosts, the ghosts have unboundedly 
negative energy density and lead to the absence of 
a stable vacuum state. In a Lorentz invariant theory 
this instability is instant as the decay rate of the 
vacuum is infinity. It is suggested that if there is 
a Lorentz non-invariant cut-off in the theory and 
the cut-off scale is enough low, it is possible 
to keep the instability at unobservable level \cite{Cli2}. 
In the DGP model, the strong coupling scale $\Lambda_c$ 
may serve as the cut-off scale. It is needed to calculate the 
decay rate of the vacuum and to see whether the 
self-accelerating universe can survive beyond the age 
of our Universe. It is also necessary to check the 
validity of the linearized analysis \cite{Gab, Dva3, Def5, 
Dva4, Koy3}. Several non-perturbative solutions 
indicate that the self-accelerating universe would 
be unstable \cite{Dva4, Gab}. Then we are naturally 
lead to ask what does the solution decay to 
\cite{Izu2}. This is still an open question.  
See \cite{Koy5} for a review on the issue of the ghost
in the DGP model.

Finally, it was pointed out that time-dependent 
perturbations around the spherically symmetric 
spacetime have a sound speed greater than 1 \cite{Adm}. 
Again there are debates whether this is a problem 
or not. One subtlety is that this argument is
based on the effective theory for the scalar mode
and it is not clear this effective theory captures 
the property of the full gravitational perturbations in the model
\cite{Gab2}.
In addition, causality should be defined in 
a 5D spacetime and it is not clear that 
the super-luminality in the 4D effective theory 
really means the breakdown of causality in 
the full 5D theory. 

Although it is still not clear whether 
we should deny the DGP model as a consistent 
theory due to these problems, this certainly 
demonstrates the difficulty for the large 
distance modification of gravity to explain 
the late time accelerated expansion of the Universe. 
It is necessary to seek improved models that 
can avoid these problems.

\section{Conclusion}
In this article, we review the attempts to address the 
cosmological constant problem and the dark energy problem 
in braneworlds. 

The cosmological constant problem
has resisted solution for many years. The conventional 
approach relies on 4D low-energy physics. This was 
a natural way of attacking the problem as in a conventional 
KK compactification, the size of extra dimensions must be 
smaller than TeV$^{-1}$ and, below the TeV scale, our Universe
can be described by the 4D effective theory. 
However, the braneworld 
picture completely changes the notion of 
extra dimensions. The extra dimensions can be large.
For 6D spacetime, the size of the extra dimensions 
can be $L \sim 10\mu$m, with the 6D planck scale 
$10$ TeV. Then above the scale $L^{-1}$, the Universe 
is described by $6D$ and the 4D effective theory 
cannot be used. In fact the energy density for 
the cosmological constant necessary to explain 
the present accelerated expansion is roughly $\rho_{\Lambda}
\sim L^{-4}$. Moreover, the way the vacuum energy gravitates
in our 4D Universe is completely different in the 
braneworld. Again in a 6D spacetime, 
the vacuum energy on a 4D brane does not curve the 
4D spacetime but just changes the geometry of the 
extra dimensions. This leads to the self-tuning idea 
where the change of the vacuum energy in 4D spacetime
is compensated by the modification in the geometry 
of extra dimensions. Although it was shown that 
the simple non-supersymmetric model does not work, 
it is hoped that the supersymmetric version of the model
can realize the self-tuning. A close inspection
reveals many problems in this approach, but further 
studies are necessary to judge whether the self-tuning 
idea really works or not. The outstanding problem
is to know whether the 4D spacetime settles down to 
a static solution due to the self-tuning mechanism 
if there is a phase transition in the 4D spacetime. 
This requires a regularization of the branes and the 
analysis of the time dependent dynamics in the 6D spacetime. 
Based on the hope for the existence of self-tuning,
the Supersymmetric Large Extra Dimensions (SLED) model 
is proposed as a framework to address the cosmological 
constant problem and the dark energy simultaneously. 
The self-tuning mechanism is supposed to cure the 
problem of the large vacuum energy produced by 
the phase transition in the 4D spacetime. This mechanism 
relies on the supersymmetry in the 6D spacetime, but
supersymmetry is inevitably broken on a brane 
at least at TeV scale. This breakdown 
is mediated to the bulk only gravitationally 
and creates a weak potential for the radion which 
is the size of the extra dimensions. The potential 
energy is determined by the supersymmetric breaking 
scale in the bulk and it is argued that if the size of the 
extra dimensions is $10 \mu$m, the 6D Planck 
scale is $10$ TeV and the potential energy for 
the radion has  
the right amplitude to explain the present accelerated 
expansion of the Universe. The potential depends on the 
details of the spectrum of theory and it remains 
to be seen whether this proposal can work or not in 
a concrete realization of the models in string theory. 

The late-time 
accelerated expansion of the Universe is a new 
problem forced by the discovery made by astronomers
in 1998. An interesting possibility to explain this 
is a large distance modification of gravity. 
Again, the braneworld picture plays an essential 
role. The braneworld model provides 
a concrete example where gravity leaks off the 
brane and modifies the 4D GR
on the brane at large distances. The DGP model 
is the simplest model that realizes this idea. 
The action for the model is very simple. The 
5D spacetime is just a Minkowski spacetime described 
by Einstein gravity. We are living on a 4D brane 
where 4D gravity is assumed to be 
induced. Despite the simple 
set-up of the model, gravity in this model is 
remarkably complicated. In fact there exists 
a solution (the self-accelerating universe) where the 
accelerated expansion of the Universe is realized just by the 
modification of gravity. We focused on the possibility to 
distinguish this model from dark energy models in GR
by combining various observations. This leads
to an interesting possibility to find a 
failure of GR at cosmological scales. 
Although the DGP model is the simplest model 
where we can address many issues from 
a simple action, the model is not free from 
problems. In particular, it has been shown that 
there exists a ghost in a self-accelerating 
universe. It is crucial to study how we can 
avoid the decay of the self-accelerating 
universe in order for the observational 
tests of the model to make sense. 

The attempts to use higher-dimensional gravity 
and branes to address the cosmological constant 
problem and dark energy are new but 
there has been much progress. 
 In this article, we only covered several 
attempts among them. We see that these attempts 
bring us a completely new way of attacking long-standing 
and tough problems although none of the models is  
completely successful so far. We hope further developments of the 
models based on these attempts lead to solutions for
the long-standing problems.  

\section*{Acknowledgments}
KK is supported by PPARC/STFC. We would like to thank R.~Maartens
for a careful reading of this manuscript.

\end{document}